# Label- and slide-free tissue histology using 3D epi-mode quantitative phase imaging and virtual H&E staining

Tanishq Mathew Abraham[1], Paloma Casteleiro Costa[2], Caroline Filan[3], Zhe Guang[4], Zhaobin Zhang[5,6], Stewart Neill[5,7], Jeffrey J. Olson[5,6], Richard Levenson[8], Francisco E. Robles[4,*]

[1]Dept. of Biomedical Engineering, University of California, Davis, Davis, CA 95616, USA
[2]School of Electrical and Computer Engineering, Georgia Institute of Technology, Atlanta, GA 30332, USA
[3]George W. Woodruff School of Mechanical Engineering, Georgia Institute of Technology, Atlanta, GA 30332, USA
[4]Wallace H. Coulter Department of Biomedical Engineering, Georgia Institute of Technology, Atlanta, GA 30332, USA
[5]Winship Cancer Institute, Emory University, Atlanta, GA 30322, USA
[6]Department of Neurosurgery, Emory University School of Medicine, Atlanta, GA 30322, USA
[7]Department of Pathology & Laboratory Medicine, Emory University School of Medicine, Atlanta, GA 30322, USA
[8]Dept. of Pathology and Laboratory Medicine, UC Davis Health, Sacramento, CA 95817, USA
*robles@gatech.edu*

## Abstract
Histological staining of tissue biopsies, especially hematoxylin and eosin (H&E) staining, serves as the benchmark for disease diagnosis and comprehensive clinical assessment of tissue. However, the process is laborious and time-consuming, often limiting its usage in crucial applications such as surgical margin assessment. To address these challenges, we combine an emerging 3D quantitative phase imaging technology, termed quantitative oblique back illumination microscopy (qOBM), with an unsupervised generative adversarial network pipeline to map qOBM phase images of unaltered thick tissues (i.e., label- and slide-free) to virtually stained H&E-like (vH&E) images. We demonstrate that the approach achieves high-fidelity conversions to H&E with subcellular detail using fresh tissue specimens from mouse liver, rat gliosarcoma, and human gliomas. We also show that the framework directly enables additional capabilities such as H&E-like contrast for volumetric imaging. The quality and fidelity of the vH&E images are validated using both a neural network classifier trained on real H&E images and tested on virtual H&E images, and a user study with neuropathologists. Given its simple and low-cost embodiment and ability to provide real-time feedback in vivo, this deep learning-enabled qOBM approach could enable new workflows for histopathology with the potential to significantly save time, labor, and costs in cancer screening, detection, treatment guidance, and more.

## Introduction
Histopathology is the gold-standard for diagnosing disease, guidance of surgical margins during lesion resection, and overall clinical evaluation of tissue[1]. To visualize tissue architecture, labor-and-time-intensive tissue processing is currently required. During the most common histopathology procedure, an excised tissue specimen is fixed in formalin and paraffin embedded (FFPE), sectioned to generate micron-thick slices, and then mounted onto microscope slides. Those slides can then undergo a number of different staining procedures with the most common being hematoxylin and eosin (H&E) staining, in which hematoxylin stains cell nuclei purple and eosin stains the extracellular matrix, stroma, and cytoplasm pink[1]. This standard, widely used process typically takes eight hours or more to complete. Consequently, fast, real-time tissue assessment with H&E-like contrast would have the potential to improve a number of medical procedures, ranging from surgical margin assessment to cancer screening and more.

In an effort to gain real-time pathology-level tissue assessments for use in surgery and other clinical fields, alternate microscopy techniques have been employed to provide imaging feedback during tissue excision. Some of these techniques include rapid tissue staining followed by linear[2,3] and nonlinear fluorescence microscopy[4]; as well as label-free approaches ranging from ultraviolet-based methods and autofluorescence[5–7] to more complex nonlinear techniques[8–11]. Many of these methods have also incorporated virtual staining pipelines to obtain images that are familiar to pathologists and thus avoid the need for further training on each imaging modality[2–4,11–14]. While promising, these methods have certain potential downsides, as they variously rely on staining the imaged tissues, employ UV light, and/or use complex and expensive nonlinear methods to achieve virtual histology. Further, translation of these technologies to in-vivo applications is challenging or infeasible given the need for exogenous agents, concerns regarding tissue damage, and technological hurdles. These challenges limit the applicability of virtually stained microscopy and slide-free histology and point out the advantages of a microscopy method that could provide histopathologic information quickly (real-time), non-destructively, and with high-resolution in 3D, using simple, low-cost instrumentation.

To achieve these desired capabilities, we propose the use of virtual-H&E-stained images obtained with quantitative oblique back-illumination microscopy (qOBM) as a method of real-time histopathology for excised tissue samples, and with a clear path to future into in-vivo applications. qOBM is a label- and fixative-free, wide-field, low-cost microscopy technique capable of obtaining sub-cellular resolution, quantitative phase images of thick, scattering tissue samples using same-side epi-illumination[15,16]. (Thick, scattering samples refers to, for example, excised tissues without sectioning or intact organs such as brains, which cannot be imaged with transmission microscopes.) The level of 3D cellular and subcellular structural detail provided by this technology is comparable to that provided by label-free nonlinear microcopy methods, but with an embodiment that is simple and orders of magnitude cheaper, as it uses LEDs instead of femto-/pico-second lasers, is faster (wide-field vs point scanning), gentle on tissues and cells, and can be easily modified and miniaturized for in-vivo applications[17,18]. Here we advance qOBM and the field of slide-free histology by introducing an image translation method by which qOBM images are virtually stained to resemble H&E-stained images.

The approach leverages deep learning, specifically generative adversarial networks (GANs)[19], which have been employed to generate virtual H&E histology from alternative microscopy modalities such as quantitative phase imaging[14], reflectance confocal microscopy[20] and photoacoustic remote sensing microscopy[21], among others[13]. This approach typically requires training datasets in which the alternative microscopy images can be pixel-registered with the target domain images (e.g., H&E), and most often relies on the use of thin tissue sections. Here, such pixel-registered datasets are unobtainable as qOBM imaging is performed on fresh tissue whereas ground-truth H&E images are subject to tissue distortions from histological processes. To work around the lack of one-to-one pixel matching, we turn to cycle-consistent GANs (CycleGANs)[22]. Recent reports (using fluorescently labeled tissue and/or UV light) have demonstrated the utility of such networks for virtual H&E staining while relaxing the pixel-matching constraint[23–25]. Here, we demonstrate the efficacy of CycleGANs for virtual H&E staining of qOBM images. This combination has the potential to reduce the time needed to acquire H&E images from hours or even days to ~1 second.

To demonstrate the clinical utility of this method, we primarily focus on imaging brain tissue and differentiating between healthy and tumor regions (this represents one of many potential applications). To date, identifying brain tumor margins intraoperatively remains a significant clinical challenge; thus, neurosurgeons are often conservative with excised margins to minimize damage to healthy brain tissue vital for neurological function. However, this approach can lead to incomplete resections and tumor recurrence. Novel intraoperative methods such as 5-aminolevulinic acid (5-ALA) in-vivo staining have shown promise for improving clinical outcomes[26–28], but they are not without their limitations. For example, 5-ALA exhibits variable uptake based on brain morphology[27], and has limited sensitivity for low-grade disease and infiltrative tumor cells even in high-grade tumors[29–31]. Real-time, label-free image guidance with H&E-like contrast has the potential to significantly improve neurosurgical outcomes, particularly if deployed in situ (that is, in the surgical site rather than on excised specimens).

In this study, we first demonstrate conversion of qOBM images to vH&E (i.e., qOBM-to-vH&E conversion) using mouse liver specimens, which have a simple and homogenous structure, to establish the feasibility and effectiveness of the approach, as well as to show its utility for imaging a variety of tissue types. Then, we demonstrate qOBM-to-vH&E conversion using tissues from a rat glioma tumor model and human glioma specimens. To validate the results, we (1) trained a classifier on real H&E images of tumor and healthy tissues and then tested on virtual H&E images, and (2) performed a user study with five board-certified neuropathologists. The proposed qOBM-to-vH&E conversion pipeline permits a novel histopathology workflow (**Fig. 1**) that has the potential to reduce the time and costs associated with obtaining histological H&E images. Further, the level of histological detail with H&E-like contrasts achieved by the proposed simple and label-free method is exemplary and paves the way for novel capabilities in a number of medical applications.

# Results
### Virtual staining of label-free qOBM images of fresh mouse liver

To establish the feasibility and effectiveness of unpaired image-to-image translation from qOBM to H&E, we first attempted to generate vH&E images of healthy mouse liver specimens, which demonstrate a consistent, well-defined microanatomy primarily comprising of well-organized hepatic cells and blood vessels. qOBM images of freshly excised liver tissue specimens (N=8), donated from otherwise discarded tissue, were obtained with a 60X objective (0.7 N.A., 270 x 270 µm field of view, with lateral resolution of 0.6 µm and cross-sectional/axial resolution of 2.5 µm). qOBM images, including real-time processing, were acquired at 10Hz. All animal experimental protocols were approved by Institutional Animal Care and Use Committee (IACUC) of the Georgia Institute of Technology and Emory University. Tissues were subsequently submitted for histological processing to obtain H&E slides (sections were ~5 µm thick). Prior to CycleGAN training, the qOBM images were contrast-enhanced and grayscale inverted. The images were divided into 512 x 512 pixel (~ 70 x 70 µm) tiles for training. We used a standard ResNet-based generator architecture and a PatchGAN discriminator, training on 2358 qOBM and 1737 H&E tiles for 200 epochs at a batch size of 4.

**Figure 2** shows representative results. First, the native qOBM phase images (**Fig. 2A**) show clear cellular and subcellular detail that closely parallels the structure of the traditional H&E images (**Fig. 2C**), making qualitative assessment of the translation relatively simple. For example, in qOBM, with contrast generated by the refractive index properties of the tissue, hepatocyte nuclei appear dark and possess subtle but appreciable subnuclear structure/texture (shown in insets), while red blood cells appear bright. **Figure 2B** shows the translated vH&E image, which preserves the general structure of the qOBM image, with a high-fidelity style conversion to H&E. Specifically, nuclear and subnuclear structures of the hepatocytes are converted appropriately, with the expected purple hue and texture. The network also correctly enhances and converts nuclei that can be difficult to identify in qOBM (white arrow), although some are occasionally missed (yellow arrow). The missed nuclei occur in areas near capillaries; this is likely due to the fact that the capillary structures in the qOBM images of fresh tissues are different (better preserved and continuous) than in the target H&E images of processed tissue sections in which the capillaries appear more fragmented (blue arrows highlight vessel structures). Consequently, the network may at times not appropriately deal with such structures. Training with more data could potentially resolve these small errors; nevertheless, the overall structure of the tissue is well preserved and is consistent with the appearance of healthy mouse liver.

We also note that structures observed in qOBM that are not present in H&E, such as small bright white droplets—likely composed of lipids—are correctly ignored in the vH&E images and do not produce unwanted artifact. And red blood cells, also depicted in bright white in the phase image, are correctly translated to their characteristic bright red hue in H&E. These results confirm that CycleGANs can successfully translate qOBM quantitative phase images of thick fresh tissue to H&E-like images without needing pixel-matched paired images.

**Virtual staining of label-free microscopy images of rat brain tumor**

Having established the qualitative ability to translate qOBM images into vH&E using a relatively simple and homogeneous sample type, we next turn to the more challenging task of virtually staining complex brain tissue (healthy and tumor) and later providing quantitative metrics of translation fidelity.

qOBM imaging of fresh tissues from a 9L gliosarcoma rat tumor model (N = 14) was performed as described in Costa et al.[17] (also see **Materials and Methods**); this tumor was chosen because of its similarity to high-grade human gliomas. Treated animals had tumors confined to one hemisphere, leaving the other as control. Two healthy mice were also imaged and analyzed as additional controls (thus a total of N=16 animals were analyzed). Images were acquired with a 60X, 0.7NA objective. During the imaging sessions, the brains were scanned laterally and axially (volumetrically) in an automated manner to acquire data from different regions of the brain. Following qOBM imaging, the brains were formalin fixed, embedded in paraffin wax, cut into thin (5 μm) sections, and stained with H&E.

Four general tissue subtypes were observed and characterized with qOBM in the 9L gliosarcoma model. **Figures 3A, B** show two densely hypercellular tumor regions, one with a malignant sarcomatous population (**Fig. 3A**), and another with a malignant glial component (**Fig. 3B**). This biphasic tumor tissue pattern is characteristic of gliosarcomas. Additionally, **Fig. 3C** demonstrates healthy basal ganglia (with the presence of white matter bundles) and **Fig. 3D** shows healthy cortex. We trained a single CycleGAN model for qOBM-to-vH&E conversion on an image set representing all four subtypes with a total of 1377 qOBM and 1744 H&E tiles of size 512 x 512 pixels, trained for 200 epochs at a batch size of 4. Qualitatively, the CycleGAN provides qOBM-to-vH&E conversions (**Fig. 3E, F, G, and H**) that are remarkably similar to standard H&E, provided for comparison in **Fig. 3I, J, K, and L**. For instance, **Fig. 3E** (vH&E) clearly shows the same overall pleomorphic, herringbone-shaped spindle cell structure shown in **Fig. 3I** (real H&E), and **Fig. 3F** shows hyperchromatic appearance (dark purple color) of the tumor cells. In the basal ganglia, the vH&E image clearly shows the eosinophilic (deep pink color) white matter bundles, consistent with the real H&E image (**Fig. 3K**). Finally, cortex regions such as normal basal ganglia exhibit the appropriate cellularity, with blood cells (with large phase values) correctly translated to an intense red hue.

**Figure 4** shows qOBM-to-vH&E conversion of specimens that were not shown to the network during training, consisting of an admixture of healthy brain and tumor. The conversion is successful and shows excellent agreement with the style and appearance of real H&E images. Specifically, the examples in **Fig. 4** show clear lines of delineation between the tumor and brain tissue, and a mesenchymal transition characteristic of the 9L gliosarcoma rat model. The non-tumor brain tissue also demonstrates reactive characteristics such as high cellularity as expected of tissue adjacent to tumor. These results highlight the ability of the qOBM-to-vH&E conversion network to make correct inferences even when presented with structures outside of those explicitly provided in training. We attribute this capability, in part, to the close resemblance of the native qOBM phase images to histology, where again the style/mode difference between qOBM to H&E is relatively minor (particularly when compared to other label-free 3D scattering-based imaging technologies[20,32]).

To quantitatively evaluate the qOBM-to-vH&E conversion, we trained a convolutional neural network classifier to discriminate between H&E images of healthy and tumor tissue and observed its performance on vH&E images. The tissue class (i.e., ground truth healthy vs. tumor) for each qOBM and H&E image was known a priori based on the anatomical location of the implanted tumor cells. The classifier was trained with 5-fold cross validation on 1395 real H&E tiles to discriminate between healthy and tumor, which yielded an accuracy of 99.4 ± 0.8% on a held-out test set of 349 512 x 512-pixel tiles. The classifier was then employed on 270 vH&E tiles generated by the CycleGAN and displayed an accuracy of 95.2 ± 2.8% (**Fig. 5**). This suggests that the translated images preserve both the style and diagnostic information content of the traditional H&E images.

**Virtual H&E staining of mosaics and tomographic volumes**

The qOBM system used in these studies was equipped with lateral and axial automated stages that enable scanning tissue in all directions to create large mosaics, as well as tomographic volumetric datasets. **Figure 6** demonstrates a virtual H&E strip mosaic (6.3 mm x 270 µm) of a rat brain, while **Fig. 7** and **Supplementary Videos 1,2** show a vH&E 3D rendered volume (270 µm x 270 µm x 60 µm) of a rat brain tumor margin. In **Fig. 6**, the overall margin delineation between tumor tissue and normal tissue based on the cellularity is clearly apparent, with excellent agreement to H&E. **Figure 7** demonstrates a transition from a glial tumor subtype surrounded by basal ganglia tissue structures into the sarcomatous tumor subtype, which is clearly apparent in the vH&E images. Here the robustness of the vH&E translation is evident and demonstrates a consistent color and structure in the reconstructed images stitched or stacked together using a standard process (see **Materials and Methods**), with no special consideration for the mosaic or volumetric nature of the datasets.

**Virtual staining of label-free microscopy images of human glioma specimens**

To demonstrate the potential clinical utility of the approach, the CycleGAN deep learning pipeline was employed to virtually stain qOBM images of human astrocytoma specimens. Samples consisted of freshly excised human brain tumor and tumor-edge regions of infiltrating grade 2 and grade 3 astrocytoma specimens discarded from neurosurgery. Five patient samples were analyzed. All tissues were imaged fresh within 6 hours of removal, and no modifications were made to the tissues prior to the qOBM imaging process. It is important to note that the margins of these types of infiltrating tumors, especially grade 2 astrocytomas, are extremely difficulty to identify intraoperatively, particularly in vivo where existing assessment tools lack sensitivity. All human samples were de-identified and obtained through the Winship Cancer Institute of Emory University using approved protocols.

We continued training the CycleGAN developed for rat brain tissue on an additional 837 qOBM and 372 H&E tiles of human glioma tissue. This process is often referred to as transfer learning or fine-tuning[33]. Qualitatively, this fine-tuned model performed significantly better on human specimens than when we attempted to apply a neural network trained exclusively on rat specimens (**Supplementary Fig. 5**). We also compared our fine-tuned model to training from scratch on the human glioma images alone (**Supplementary Fig. 5E, F**), observing that the fine-tuned model demonstrates significantly better subnuclear detail, especially when trained with higher resolution H&E images (**Fig. 8**).

**Figure 8A-F** show two human grade 3 astrocytomas, clearly identifiable due to their hypercellular and hyperchromatic tumor cells. In the qOBM phase images, the cells are tightly packed and display rough intranuclear texture; these are appropriately translated in the vH&E image. In **Fig. 8G-I**, we see another hypercellular human grade 3 astrocytoma. Both the virtual and real H&E show atypically shaped cells and nuclei that are an important indicator of tumor presence. Note that the qOBM image (**Fig. 8G**) contains small bright white dots throughout the image which we have exclusively observed in brain samples from patients who have received prior radiation treatments (data from a parallel study[34]). These features are only visible in the qOBM images of fresh tissues and vanish after FFPE H&E processing. Interestingly, the digital conversion to vH&E also suppresses the appearance of these structures. This is similar to the results presented in **Fig. 2**, where the lipid-like structures present in the qOBM images of liver are not displayed in the corresponding vH&E image as they are absent in the target domain H&E images. **Figure 8J-L**, present a human grade 2 (low-grade) astrocytoma. Here we observe, in both the virtual and real H&E, moderate cellularity and nuclear pleomorphism. This shows the potential of the proposed method to correctly capture H&E-like histological detail indicative of low-grade disease, which again, is extremely difficult to identify intraoperatively with existing intraoperative tools. Finally, **Fig. 8M-O**, presents a healthy human tissue specimen from the edge of a grade 3 astrocytoma tumor, where the vH&E image resembles the real H&E image, with both showing regularly shaped cell nuclei without hyperchromasia and at the expected density for normal tissue.

Volumetric stacks of human gliomas specimens can also be obtained and virtually stained, allowing us to gain additional insight about the specimen. For example, **Fig. 9** and **Supplementary Videos 3,4** show a volume of a human grade 3 astrocytoma where the first most shallow image exhibits a structure consistent with normal brain tissue with the exception of a single atypical cell (as indicated by the arrow in **Fig. 9C** at a depth of $Z = 2$ µm). These characteristics alone would not be sufficient to diagnose as tumor or warrant excision of the tissue if seen in-vivo intraoperatively. However, as we image deeper into the sample, the tissue exhibits higher cellularity with larger, hyperchromatic cells becoming evident, reflecting the presence of tumor. By being able to move axially (deeper) into the tissue, we can gain additional information, including seeing increased counts of more irregular nuclei, which indicates tumor.

**Neuropathologist validation of virtually stained qOBM images**

To further validate the potential clinical utility of the virtually stained qOBM images, we performed a user study with American Board of Pathology certified neuropathologists. We collated a set of 30 vH&E images of the rat brain tumor model and 20 vH&E images of human gliomas along with corresponding real H&E images, giving a total of 100 images. These images were reviewed by five neuropathologists, who were asked to respond to 3 questions: (1) Are tumor cells present in the image? (Y/N/Cannot assess); (2) If this field of view were representative of a larger region, would you recommend continued resection? (Y/N); and (3) How confident are you in this evaluation? (1, unsure to 5, very confident).

To assess accuracy, we designated the following criteria: For the H&E and vH&E images of the animal model, ground truth was based on a-priori knowledge of the location of the tumor (see **Methods and Materials**). For the human H&E images, ground truth was taken to be the consensus answer from the five neuropathologists. For the vH&E images, ground truth was based on the evaluation of the same specimens after H&E processing, which in this case also agreed with consensus of the vH&E images.

| Parameter | H&E | Virtual H&E | Statistical Significance |
|---|---|---|---|
| **Accuracy** | 94% | 96% | N.S. |
| **Overall Group Concordance** | 0.74 | 0.81 | - |
| **Diagnostic Confidence** | 4.6 | 4.7 | N.S. |

**Table 1 – Neuropathologist user study comparing standard H&E and virtual H&E for interpretation.** Overall accuracy of assessing tumor cell presence is reported. Group concordance is reported as average pairwise Cohen's Kappa value. Diagnostic confidence is scored from 1 (unsure) to 5 (very confident) and average score is reported.

The responses of the neuropathologists (results summarized in **Table 1**) further validate that the vH&E images and the H&E-stained tissue sectioned images are of similar quality. Both the accuracy and the quality ratings between the two modalities were high, and with no statistically significant differences, suggesting that the virtual staining method produced high-quality discernible images that would be clinically useful for interpretation by neuropathologists. Specifically, the overall accuracy for assessing the presence of tumor cells on the real H&E and vH&E images was 94% and 96%, respectively. The inter-group concordance using the average pairwise Cohen's Kappa value for recommended continued resection demonstrates a near-perfect level of concordance between the pathologists for both the H&E and virtually stained results (0.74 and 0.81 for the real and virtual H&E, respectively). Finally, the diagnostic confidence was also similar for both types of images (4.6 and 4.7 for the real and virtual H&E, respectively).

This survey supports the effectiveness of qOBM-to-vH&E conversion for clinical applications including intraoperative guidance and more.

## Discussion

Traditional biopsies require tissue excision, histological processing, and examination by a pathologist, a long process that is challenging to accomplish in a surgical environment; the logistics also affect the feasibility of many other clinical tasks such as cancer screening. For intraoperative surgical applications, rapid pathological assessments have thus far been limited because standard FFPE histology requires time-consuming (overnight or longer) tissue processing, leading to the usage of faster but technically challenging approaches such as frozen sections. Various slide-free and label-free microscopy technologies have been developed to address these problems, but those that do so successfully face significant challenges for in-vivo applications and require complex, bulky and expensive systems to achieve H&E-like images. Here we demonstrate the feasibility of qOBM imaging for rapid assessments, supplementing it with a deep-learning-based framework to obtain H&E-like results from its otherwise clinically unfamiliar grayscale phase-contrast images. To this end, we made use of an unpaired image-to-image translation algorithm known as a CycleGAN to perform a qOBM-to-virtual H&E conversion. We demonstrated this approach with both liver and brain tissue, from three species (mouse, rat and human). The converted images rendered the subcellular and cytoplasmic detail present in the original qOBM image to resemble familiar H&E contrast. The ability of qOBM to provide real-time, label-free, tomographic images of thick tissue specimens with remarkable agreement to traditional H&E histology is feasible because the style/mode difference between qOBM to H&E is relatively minor and facilitates the use of unpaired image-to-image translation. Additionally, qOBM can be implemented as a handheld probe[17,18], enabling in-vivo imaging for potential intraoperative, monitoring or screening, as well as other clinical and biomedical applications.

Previous studies have explored the use of CycleGANs for virtual H&E staining with confocal fluorescence[23], MUSE[24], and UV photoacoustic microscopy[25]. Two alternative methods were also compared for MUSE-to-H&E conversion but it was observed that best performance was obtained with CycleGANs[24]. This motivated the use of CycleGANs in our approach. Here we identified several steps that improved CycleGAN performance for qOBM-to-vH&E conversion: First, grayscale inverting the qOBM images was necessary for the success of conversion since nuclei (especially of tumor cells) have higher refractive index and thus show a higher brightness in qOBM images whereas the background is dark, opposite to how such structures appear in standard H&E (**Supplementary Note 1**). Second, transfer learning helped with the performance of human glioma qOBM-to-vH&E conversion (**Supplementary Note 2**). We also found that our models for transforming individual FOVs generalized well to volumetric stacks and stitching large fields of view, which had been a challenge in other image translation pipelines[20]. We evaluated our conversion efforts with a proxy deep learning classification task, observing that a classifier trained on standard H&E performs similarly on vH&E images. Additionally, we validated our model performance with a study involving 5 neuropathologists, who found the virtual H&E images functionally equivalent to the standard H&E images for potential surgical guidance.

Moreover, qOBM enables 3D sectioning with vH&E contrast, overcoming limitations of many current slide-free histology methods. Volumetric imaging can be especially important as it can provide a more comprehensive understanding of a tissue specimen and therefore enables more accurate diagnoses[35]. In fact, in this work we observed that the volumetric imaging capabilities of qOBM can provide critical insight for human specimen that could otherwise be missed with surface level (2D) technologies, even ex vivo. Note that while the deepest vH&E slices we show here is 60 μm from the cut surface, qOBM can achieve a penetration depth of ~120 μm with 720 nm LED illumination (data not shown). The full depth range of qOBM could potentially be used for vH&E with improvements in signal-to-noise ratio. Moreover, further improvements in the penetration depth of qOBM and hence vH&E can be achieved by using longer wavelengths extending into the near-IR.

Recent work using reflectance confocal microscopy (RCM) and deep learning also showed an ability to provide pseudo-H&E virtual staining[20]. While extremely promising, this approach is not without limitations. In contrast to the case with qOBM, RCM is generally unable to capture the same level of cellular and subcellular details, resulting from inherent differences in the object-frequency-content acquired with each method[15,16]. Consequently, the RCM to pseudo-H&E pipeline[20] requires a two-step process with "ground truth" pseudo-H&E images constructed from tissues stained with acetic acid and an analytical pseudo-H&E algorithm. The proposed pipeline using qOBM and direct conversion to H&E overcomes these limitations and enables improved histological detail with simpler instrumentation (wide field vs. point scanning, and LED lights sources vs. lasers), while achieving the same penetration depth.

In terms of computational speed, the CycleGAN takes <1 second to acquire and virtually stain a FOV using an NVIDIA A100 GPU. For eventual clinical applications, we expect such a model to be run on more modest computer units where inference time could be longer. However, we believe there are many opportunities for further optimization of the speed of model inference, either through the use of deep learning compilers that speed up the existing model, or compression/distillation approaches[36] that train a smaller, faster model that matches the performance of the original model.

While the qOBM-to-vH&E conversion algorithm serves as a useful visualization tool for clinicians to interpret qOBM images, we envision the usage of qOBM-to-vH&E conversion as part of an AI-based diagnostic and decision support pipeline. Various diagnostic AI systems have been developed for H&E-stained images with high accuracy[37,38]. In contrast, due to the limited data available for a novel technology like qOBM, it would be challenging to develop diagnostic AI systems from scratch. Instead, the qOBM images can be converted to vH&E and diagnostic pipelines developed for H&E can then be applied. A proof-of-concept example was demonstrated here by the use of a simple CNN trained on H&E images subsequently applied to the vH&E images (**Fig. 5**). The utilization of qOBM-to-vH&E conversion may allow us to leverage recent advances in computational pathology in new settings, widening the potential of qOBM imaging and slide-free histology.

While our virtual staining results are promising and vH&E images retain diagnostically relevant features, conversion is not pixel-wise perfect. As shown in **Supplementary Fig. 4**, the CycleGAN occasionally has the tendency to hallucinate nuclei, or omit them (**Fig. 2**), primarily around blood vessels. We believe this is due to inherent differences between fresh tissues imaged in qOBM and processed tissues imaged in standard brightfield H&E images, which make the unpaired image-to-image translation difficult in certain scenarios. We note that the main difficulties appear when artifacts are present in the target domain (H&E of fixed tissues) that are not observed in the original domain (qOBM of fresh tissues). However, the model does well when additional features are present in the original domain but missing in the target domain. Future work can examine unpaired image-to-image translation techniques that better ensure the content of the original image is preserved appropriately. However, the underlying challenge limiting conversion efforts is the lack of paired pixel-matched ground truth data. Specifically, the exact same cells and structures cannot be captured by both qOBM and standard brightfield H&E due to the additional tissue processing and sectioning steps involved in the latter. This challenge is what necessitated the use of unpaired image-to-image translation. Therefore, for further improvements and pixel-wise agreement, an alternative approach could be to incorporate a secondary slide-free microscopy technology that provides images similar to H&E in a multimodal system.

Given the lack of pixel-wise ground truth, we validated the virtual H&E brain images by conducting a neuropathologist study, which indicated no significant difference between how board-certified neuropathologists interpret standard brightfield H&E and vH&E images. Future work will focus on imaging in vivo and in real-time, to be evaluated using a handheld probe to collect and virtually stain images.

The proposed technology has the potential to significantly save time, labor and expense, while enabling new capabilities for non-invasive, in-vivo imaging. For analysis of ex-vivo samples, as demonstrated here, an existing digital brightfield microscope can be modified to deliver 3D quantitative phase imaging and vH&E with qOBM for less than $500 USD. No reagents for staining are required, as this is a label-free technology. Further, as we have previously shown[17,18], qOBM can be configured as a handheld probe or endoscope which could enable novel in-vivo capabilities.

In this study, we specifically focused on the application of brain tumor margin assessment, where real-time, label-free in-vivo histological analysis is gravely needed; however, the proposed workflow enabled by deep learning-based virtual staining of qOBM images could be transformative and widely used to improve cancer screening, detection, treatment guidance, and more.

## Methods
### Label-free qOBM imaging

The qOBM system consists of a conventional inverted microscope with a modified epi-illumination scheme, as shown in **Fig. 1B**. The illumination consists of four LED light sources (720 nm) coupled into 1-mm multimode fiber optics with a 0.5 NA. The fibers are evenly distributed around the microscope objective (Nikon Plan Fluor ELWD, 60x, 0.7 NA) at a 45-degree angle from the optical axis. LEDs illuminate samples sequentially, and for each illumination, a raw bright field image is collected. By way of multiple scattering, this illumination configuration produces an effective oblique illumination[15,39]. Upon subtraction of two captures with diametrically opposed illumination, we obtain a differential phase contrast (DPC) image, $I_{DPC}$, which provides tomographic cross-sectioning capabilities with qualitative differential phase contrast.

To reconstruct a 3D quantitative phase image with qOBM, two DPC images from orthogonal angles (or shear directions) are processed and deconvoluted with the system's optical transfer function through a Tikhonov regularized deconvolution following:

$$\phi = \mathcal{F}^{-1}\left\{\frac{\sum_k \bar{I}_{DPC}^k \cdot C_{DPC}^*}{\sum_k |C_{DPC}|^2 + \alpha}\right\}$$

Here, $\phi$ represents the quantitative phase, $\bar{I}_{DPC}^k$ is each DPC image along the k$^{th}$ shear direction, alpha is a regularization parameter, and $C_{DPC}$ is the optical transfer function of the system, which can be obtained by characterizing the distribution of the multiple-scattered light passing through the focal plane within the sample[15,16].

The qOBM images capture the quantitative phase of the samples, which is directly correlated to the refractive index and dry mass of the sample. Additionally, the qOBM images show outstanding detail in all directions of illumination, have diffraction limited resolution (~0.6 μm), and a sensitivity of ~2 nm[15,17]. qOBM image acquisition is at 10 Hz (limited by the frame rate of the camera) and processing of the quantitative phase images is achieved in real-time using a regular table-top computer.

**Sample preparation and imaging**

In this work, we studied the virtual staining of qOBM images from 3 types of tissues: mouse liver, rat brain 9L gliosarcoma tumor model, and human brain tumors. All animal tissue excision and imaging protocols were approved by Institutional Animal Care and Use Committee of the Georgia Institute of Technology. All human samples were de-identified and obtained through the Winship Cancer Institute of Emory University using approved protocols. Tissues were imaged fresh and untreated, ex-vivo within 6 to 12 hours of removal. The imaged mouse livers from 8 healthy animals were donated by the Haider lab at Georgia Tech and Emory University from mice sacrificed for various purposes. The livers were excised and imaged unfixed within 3 hours of the procedure. Details about the 9L gliosarcoma rat tumor model protocol and imaging may be found in Costa et al.[17] In short, 14 Fisher rats were intracranially implanted with 9L gliosarcoma cells. The animals were sacrificed 9-12 days after the implant, and brains were excised, cut coronally to expose the tumor, and imaged unfixed within 12 hours of extraction. In this animal model, the tumor is confined to the side of the brain where the tumor cells were implanted, leaving the counter-lateral side of all treated brains as an additional control. This also allows for a-priori knowledge of the location of the tumor. Human tissue specimens from 5 patients were imaged post-surgery, within 6 hours of resection.

The qOBM imaging sessions consisted of multiple lateral and axial scans of different regions of each tissue. These scans were performed in an automated manner, enabled by the X-Y-Z automatic stages built into the microscope. Axial stacks were taken by translating the objective by 1-μm steps. The lateral scanning was performed with an overlap of 20% to facilitate stitching of mosaics and combined with axial scans.

After imaged with qOBM, all tissues were formalin-fixed for 48 hours, processed, and embedded in paraffin. Then, the samples were sliced into 5-μm sections and stained with H&E. The whole H&E sample slides were then digitally scanned by an Olympus NanoZoomer whole-slide scanner at either 20x or 40x magnification. Finally, the H&E slide scans were inspected to select similar regions to those acquired with qOBM for the CycleGAN training process, described below.

**Virtual H&E staining with CycleGAN**

We define two image domains, one for qOBM images ($X$), and one for H&E images ($Y$). We attempt to determine a transformation $G: X \rightarrow Y$. In the CycleGAN framework used here[22], there are two tasks. One task is to learn $G_X: X \rightarrow Y$ that maps $x \in X$ to $y \in Y$. The auxiliary task is to learn a generator $G_Y: Y \rightarrow X$. Additionally, we have adversarial discriminators $D_X$ and $D_Y$. $D_X$ discriminates between the fake outputs of $G_X$ and real images from the domain $Y$. On the other hand, $D_Y$ discriminates between the fake outputs of $G_Y$ and real images from the domain $X$. The CycleGAN framework then exploits the cycle-consistency property that $G_Y(G_X(x)) \approx x$ and $G_X(G_Y(y)) \approx y$. This is expressed as the following loss:

$$\mathcal{L}_{cycle}(G_X, G_Y) = \mathbb{E}_{x \sim p_{data}(x)}\left[\left\|G_Y(G_X(x)) - x\right\|_1\right] + \mathbb{E}_{y \sim p_{data}(y)}\left[\left\|G_X(G_Y(y)) - y\right\|_1\right]$$

where $\|\cdot\|_1$ is the L1 norm. This is trained with traditional least-squares adversarial losses:

$$\mathcal{L}_G(D, G, X, Y) = \mathbb{E}_{x \sim p_{data}(x)}\left[\left(D(G(x)) - 1\right)^2\right]$$

$$\mathcal{L}_D(D, G, X, Y) = \frac{1}{2}\mathbb{E}_{y \sim p_{data}(y)}[(D(y) - 1)^2] + \frac{1}{2}\mathbb{E}_{x \sim p_{data}(x)}\left[\left(D(G(x))\right)^2\right]$$

Finally, for regularization, an identity constraint is imposed:

$$\mathcal{L}_{idt}(G_X, G_Y) = \mathbb{E}_{x \sim p_{data}(x)}[\|G_Y(x) - x\|_1] + \mathbb{E}_{y \sim p_{data}(y)}[\|G_X(y) - y\|_1]$$

Thus, the full objective is:

$$\min_G \mathcal{L}_{full} = \lambda_{cyc}\mathcal{L}_{cycle}(G_X, G_Y) + \mathcal{L}_G(D_Y, G_X, X, Y) + \mathcal{L}_G(D_X, G_Y, X, Y) + \lambda_{idt}\mathcal{L}_{idt}(G_X, G_Y)$$

$$\min_D \mathcal{L}_{full} = \mathcal{L}_D(D_Y, G_X, X, Y) + \mathcal{L}_D(D_X, G_Y, X, Y)$$

where $\lambda_{cyc} = 10$ controls the impact of the cycle-consistency loss, and $\lambda_{idt} = 0.5$ controls the impact of the identity loss.

The generator architecture ($G_X$, $G_Y$) was a ResNet-based fully convolutional network described in Zhu et al.[22] Unless otherwise specified, the generator had nine residual blocks. A 70 x 70 PatchGAN[40] was used for the discriminator ($D_X$, $D_Y$). Unless otherwise specified, the discriminator had three layers. The same loss function and optimizer as described in the original paper[22] was used. The learning rate (LR) was fixed at 2e-4 the first 100 epochs and linearly decayed to zero in the next 100 epochs. A batch size of 4 was used.

qOBM images were center-cropped to 1536 x 1536 pixel images and divided into nine 512 x 512 tiles. Unless otherwise noted, all qOBM images were contrast-inverted. The H&E images were upscaled with bilinear interpolation by a factor of either 1.5x or 2x (depending on the dataset) such that the images had features of comparable pixel dimensions to those in the qOBM images.

To enable a scalable inference pipeline, we utilized a tiled inference procedure as described in Abraham et al.[24] Briefly summarized: the model was applied to overlapping 512 x 512 tiles of the original FOV and the tiles were stitched by defining a given pixel's intensity as the weighted average of intensity values from the vH&E patches which overlapped at the given pixel location. The weighting was based on a Gaussian kernel.

Since four raw captures are taken with qOBM, from which two DPC images are reconstructed, single capture- and DPC-to-vH&E conversion was also performed and compared to qOBM-to-vH&E (the images used for training come from the exact same fields of view), except the images were not inverted, since the nuclei appeared dark and therefore should be the best-case scenario for conversion efforts. Neither the raw capture nor the DPC images alone supported high-quality CycleGAN conversions (**Supplementary Fig. 3**).

For conversion of rat brain qOBM images, a single larger model was trained on all four observed tissue subtypes simultaneously. As commonly noted with CycleGANs, model size played a role in conversion performance (see **Supplementary Note 2**). Our larger model had twelve residual blocks in the generator and six layers in the discriminator. Fine-tuning of the rat CycleGAN on the human specimens simply consisted of initializing the model with the rat CycleGAN model weights and training at an LR of 2e-5. When using the high-resolution H&E for fine-tuning instead, the usual LR of 2e-4 was applied. For conversion of the qOBM strip, the full stitched strip was taken and passed into our tiled inference algorithm, rather than the individual FOVs from the strip.

**Quantitative evaluation of virtual H&E staining results**

We first trained a convolutional neural network on standard brightfield H&E images to classify between healthy (cortex or basal ganglia) regions and tumor regions. We performed five-fold cross validation. The model was trained on a total of 1744 standard H&E images, so in each fold, this led to a train-validation split of 1395 - 349 image tiles. An ImageNet-pretrained ResNet18[41] was fine-tuned with a batch size of 128 for 4 epochs. In the first epoch, only the linear head layer was trainable, and for the remaining epochs the model weights were frozen (not updatable). It was trained with a LR of 1e-2 with a short LR warmup followed by a cosine decay. The remaining three epochs were trained with all layers updatable, with a base LR of 5e-3, but using discriminative LRs[42] where early layers in the neural network have even lower LRs. These three remaining epochs were trained with a one-cycle LR schedule[43]. The mean and standard deviation of the accuracies for the classifiers trained on each of the five folds were reported.

Once accurate H&E healthy vs. tumor classifiers were trained, they were applied to vH&E images. The accuracy was calculated by comparing the labels predicted by the classifier to the ground-truth labels of the original qOBM images, and the mean and standard deviation of the accuracies were reported.

**Computational Hardware and Software**

All deep learning models were trained on NVIDIA A100 80GB GPUs. The PyTorch (version 1.9.1)[44], fastai (version 2.6.3)[45], and UPIT (version 0.2.3)[46] libraries were used for training and inference of all models.

**Clinical validation of vH&E images of brain tissue**

To evaluate the quality and usefulness of the virtually stained qOBM images compared to the gold standard H&E-stained images, we conducted a panel study with 5 board-certified neuropathologists. In this study, the neuropathologists were asked to evaluate a total of 100 180 µm x 180 µm images. The image set contained 30 real H&E rat brain images (10 tumor, 10 healthy, and 10 mixed fields of tumor and healthy), 30 virtually stained qOBM images (10 tumor, 10 healthy, and 10 mixed fields of tumor and healthy), 20 real H&E human brain tumor images, and 20 virtually stained qOBM human brain tumor images. The order of images presented in the survey was randomized with healthy and tumor regions from both humans and rats combined. For each image, neuropathologists were asked if tumor cells were present in the image (Y/N/cannot assess), based on the image if they would recommend continued resection of the area (Y/N), and how confident they were in giving that recommendation with numerical scores (1–unsure to 5–very confident).

## Figures

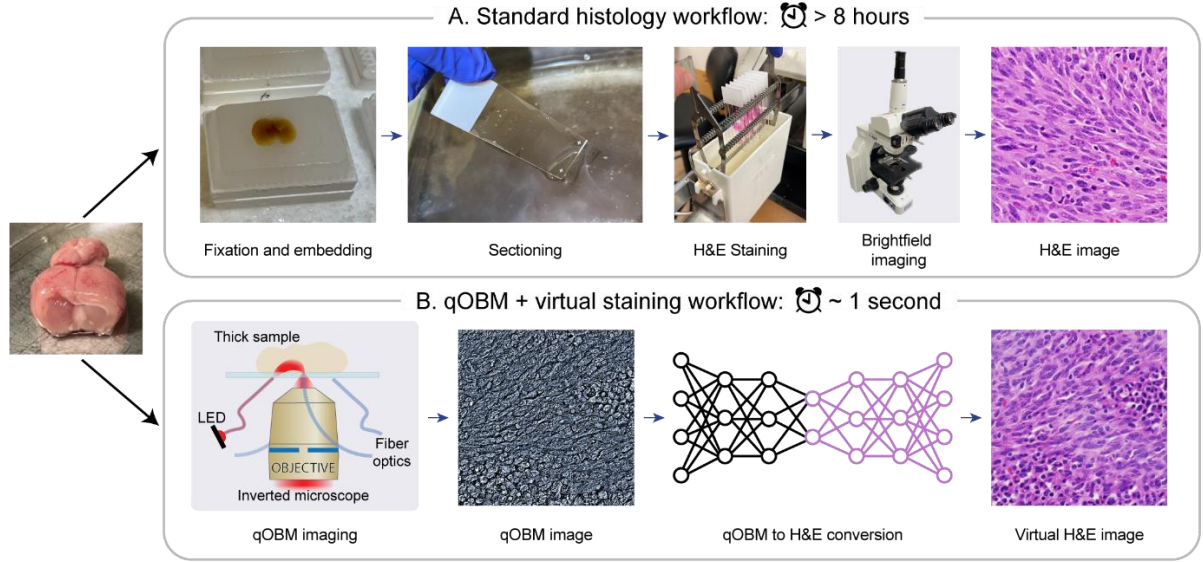

**Figure 1 - Deep learning-enabled qOBM imaging workflow.** **(A)** The standard histology workflow requires several sample preparation steps before viewing under a brightfield microscope and interpretation. This process can take about 8 hours or longer. **(B)** Our proposed workflow utilizes qOBM imaging to image a fresh specimen of tissue and virtual staining to obtain similarly interpretable images in about 1 second.

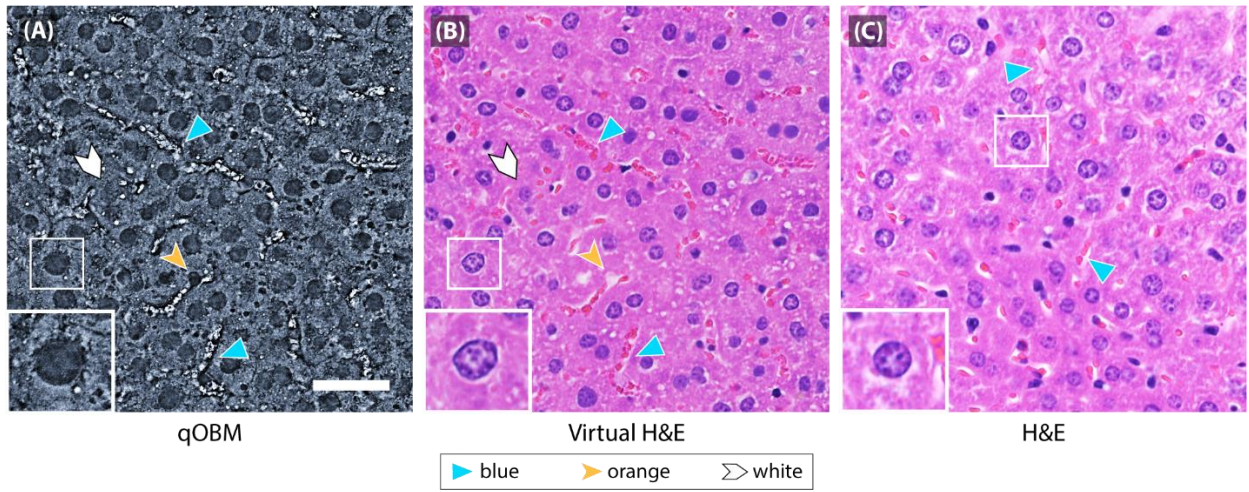

**Figure 2 - qOBM-to-vH&E conversion of mouse liver tissue.** **(A)** Label-free 60x qOBM image of mouse liver tissue. **(B)** Corresponding vH&E image. **(C)** Standard brightfield H&E image provided for comparison. The white boxes and insets show a representative appropriately converted hepatocyte with appreciable subnuclear detail, the yellow arrows refer to nuclei missed by the conversion, and the blue arrows refer to capillaries. Scale bar is 50 μm.

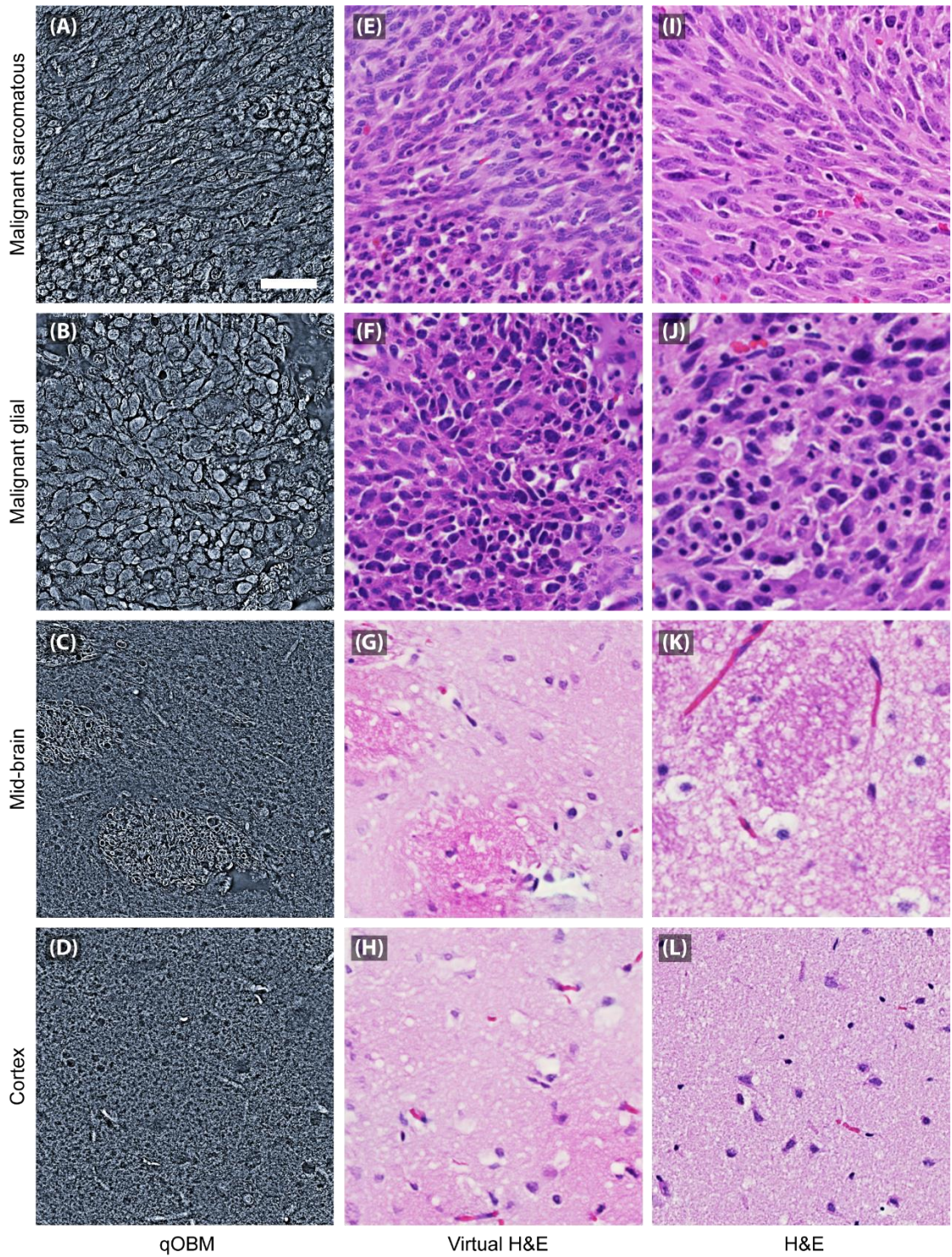

**Figure 3 - qOBM-to-vH&E conversion of brain tissue from the 9L gliosarcoma rat tumor model.** **(A)-(D)** Label-free 60x qOBM images of each of the four rat brain tissue subtypes, including two types of tumor structure (A and B), heathy basal ganglia (C) and healthy cortex (D). **(E)-(H)** Corresponding vH&E images produced by a CycleGAN trained on rat brain images. **(I)-(L)** Standard brightfield H&E images of the same tissue subtypes, provided for comparison. Scale bar is 50 µm.

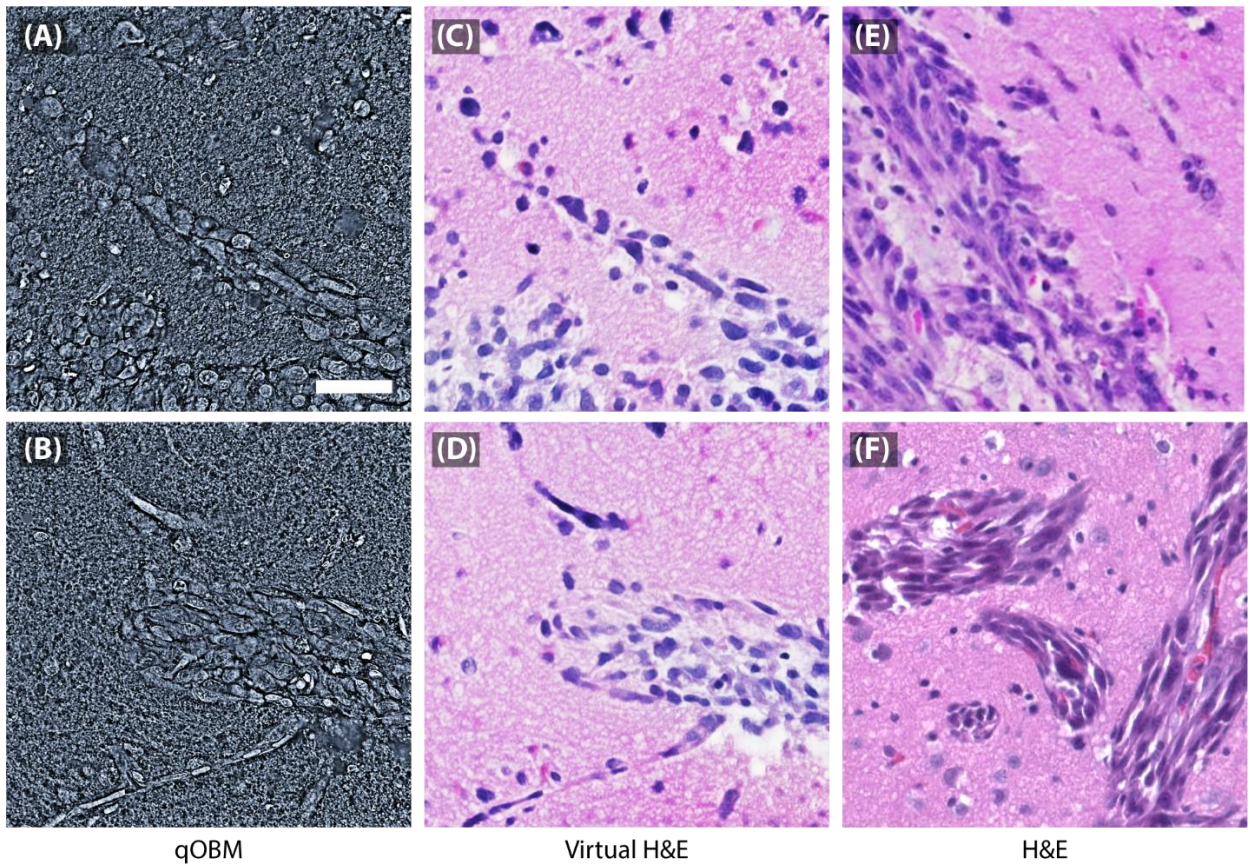

**Figure 4 - qOBM-to-vH&E conversion for images with a mix of healthy and tumor rat brain tissue, never seen during training.** **(A)-(B)** Label-free 60x qOBM images of mixed rat brain tissue. **(C)-(D)** The corresponding vH&E images. **(E)-(F)** Standard brightfield H&E images of the same tissue subtypes, provided for comparison. Scale bar is 50 µm.

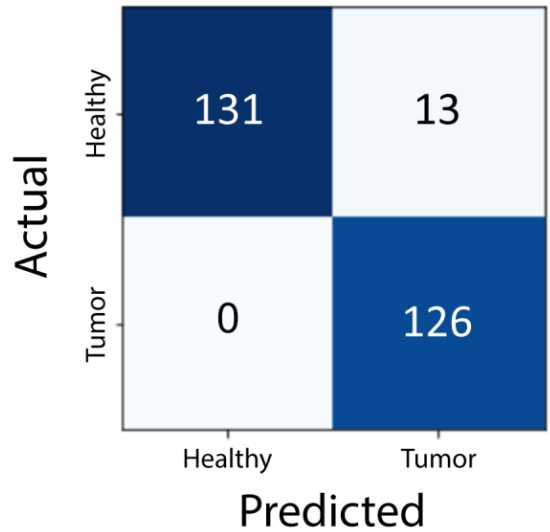

**Figure 5 – Quantitative evaluation of qOBM-to-vH&E conversion for rat brain tissue.** A classifier trained on standard H&E images to differentiate between tumor and healthy images is assessed using vH&E images. Summary of results: **(A)** accuracy of training H&E set with 5-fold cross-validation and the accuracy of the vH&E test set. **(B)** A confusion matrix of this H&E healthy/tumor classifier applied to the vH&E images.

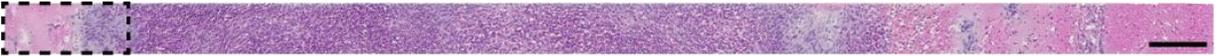

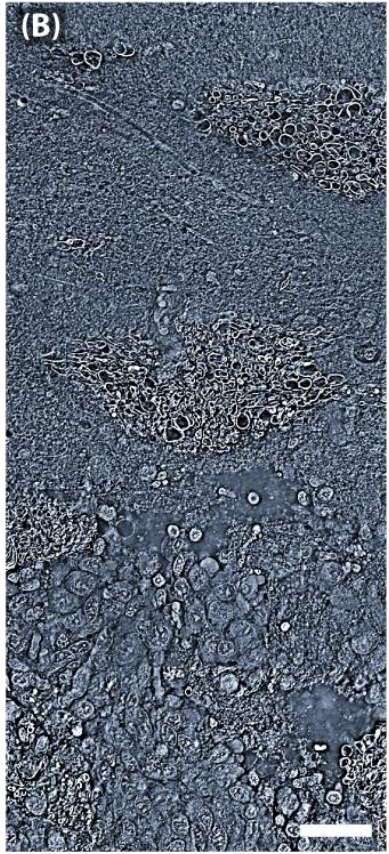
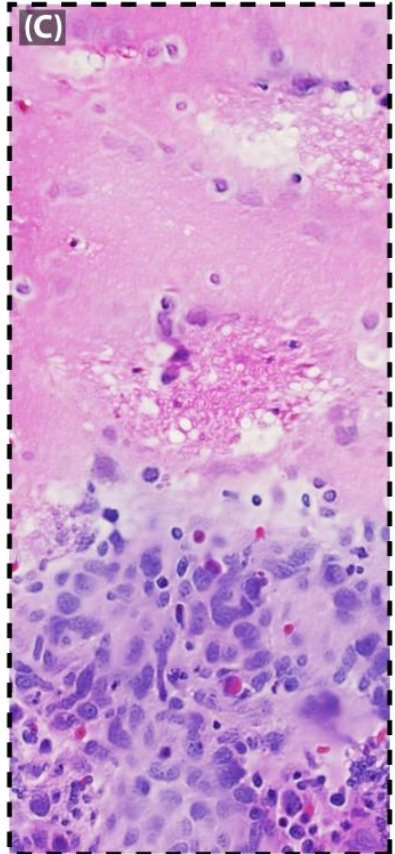
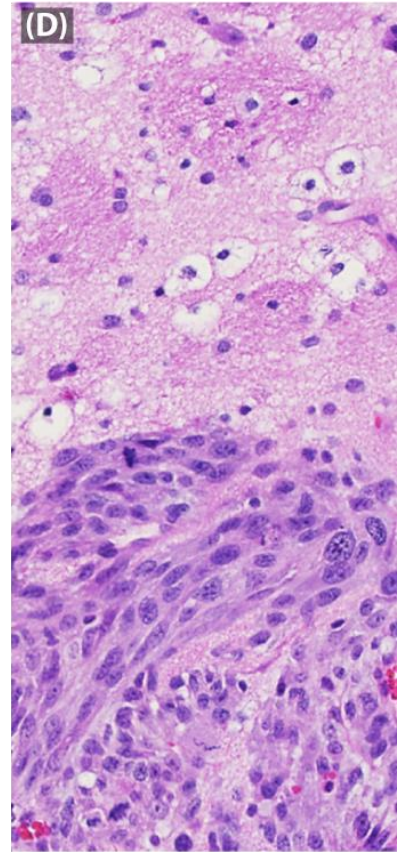

qOBM　　　　　　　　Virtual H&E　　　　　　　　H&E

**Figure 6 - Label-free qOBM imaging of a strip of rat brain tissue and corresponding vH&E.** A virtual H&E strip mosaic of rat brain tumor obtained by applying the CycleGAN trained on rat brain to the whole mosaic at once. **(A)** Virtual-H&E mosaic (6.3 mm x 270 µm). Scale bar is 300 µm. **(B)** A zoomed-in region of the label-free 60x qOBM strip (600 µm X 270 µm). Scale bar is 50 µm. **(C)** A zoomed-in region of the corresponding vH&E strip. **(D)** A zoomed in region of a brightfield H&E strip region provided for comparison.

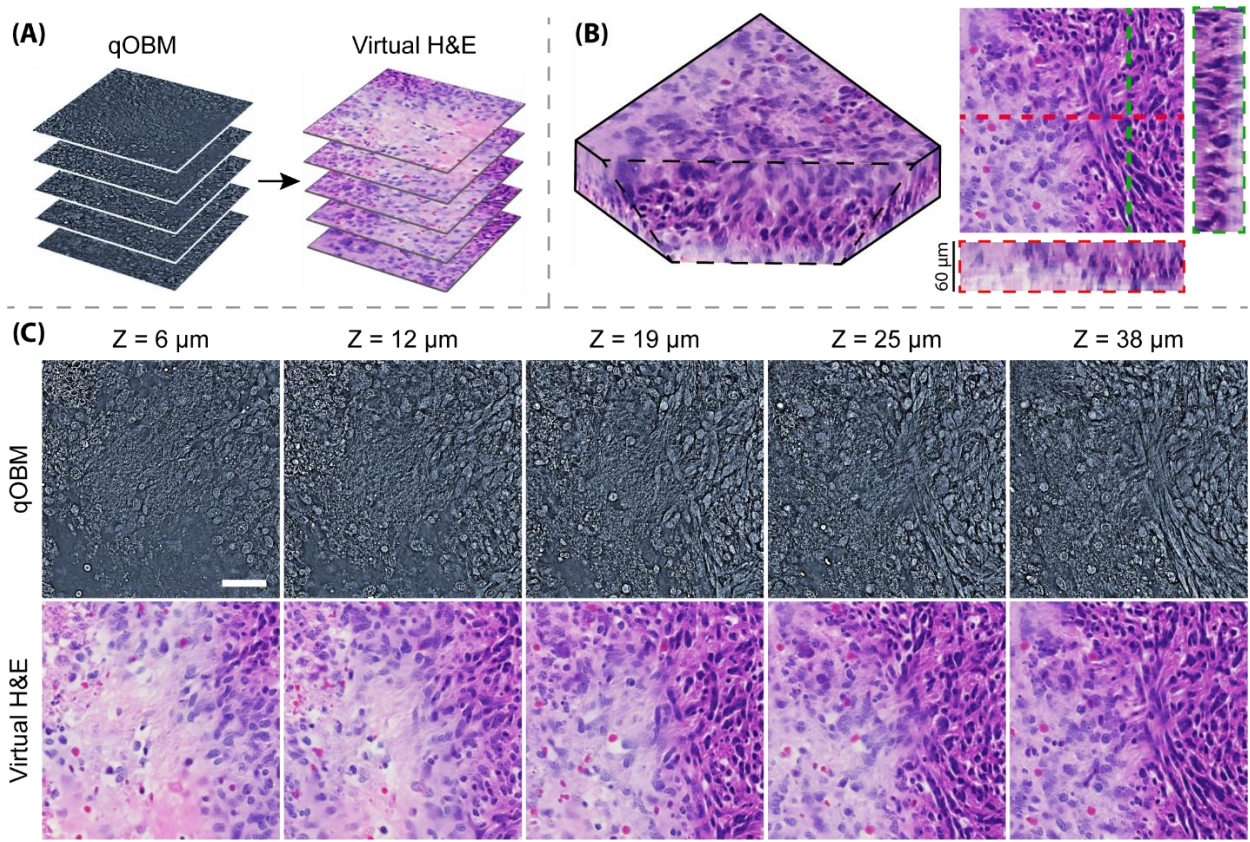

**Figure 7 - A qOBM and corresponding vH&E 3D volumetric stack of a rat brain tumor margin.** A virtual-H&E volumetric stack (270 µm x 270 µm x 60 µm) obtained by applying the trained rat brain CycleGAN to each image in the stack. **(A)** qOBM-to-vH&E conversion of the volumetric stack is depicted. **(B)** vH&E volume, with X-Y, X-Z, and Y-Z cross sections shown. **(C)** qOBM image slices at various depths and the corresponding vH&E image slices. Scale bar is 50 µm. See **Supplementary Videos 1,2** for a depth sweep-through of this volumetric stack.

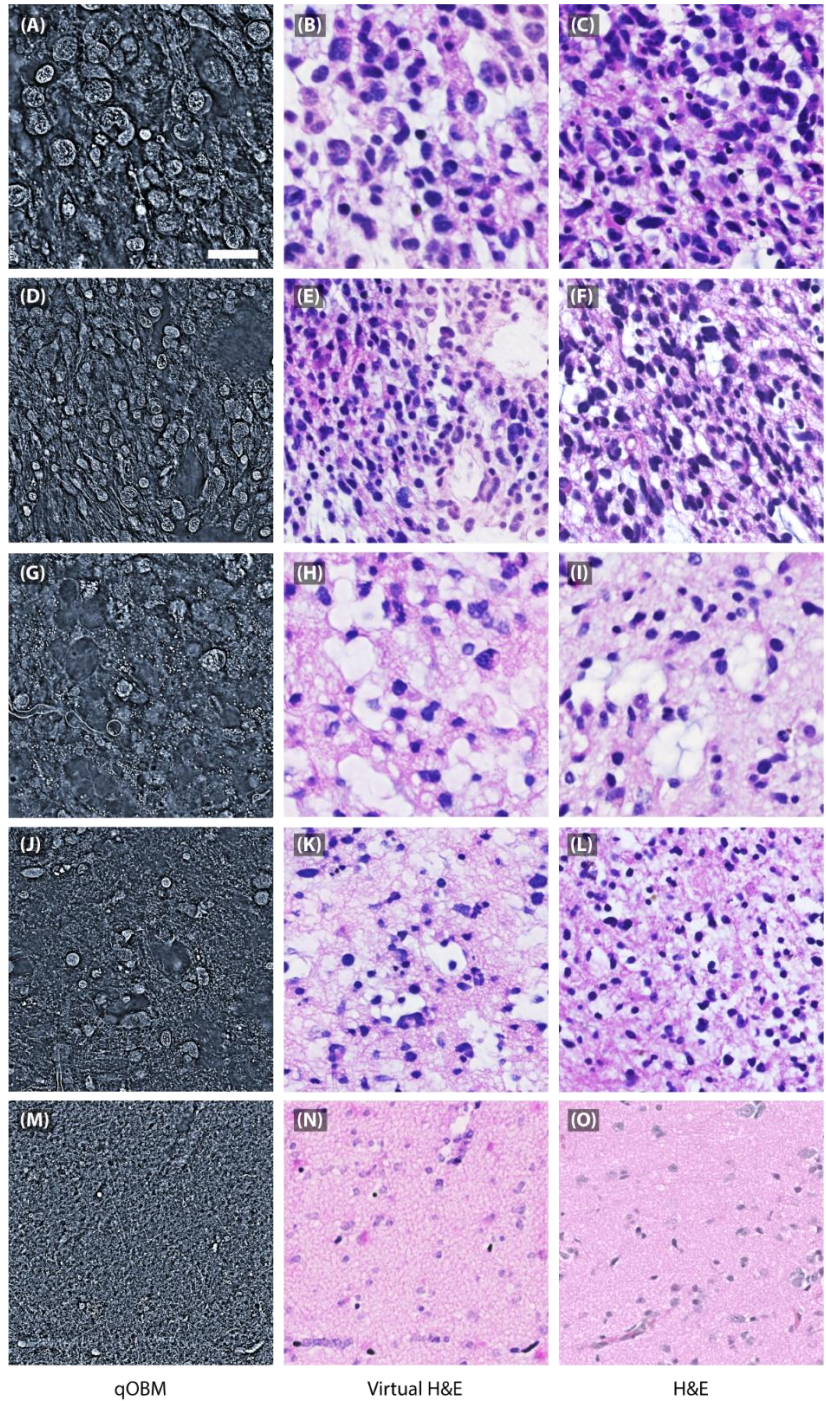

**Figure 8 - qOBM-to-vH&E conversion of human gliomas.** Each row contains a qOBM image, corresponding vH&E, and standard brightfield H&E images, provided for comparison. **(A)-(I)** Three separate human grade 3 glioma specimens (one per row). **(J)-(L)** Human grade-2 (low-grade) glioma specimen. **(M)-(O)** Healthy human tissue specimen from the edge of a grade 3 astrocytoma. Scale bar is 50 µm.

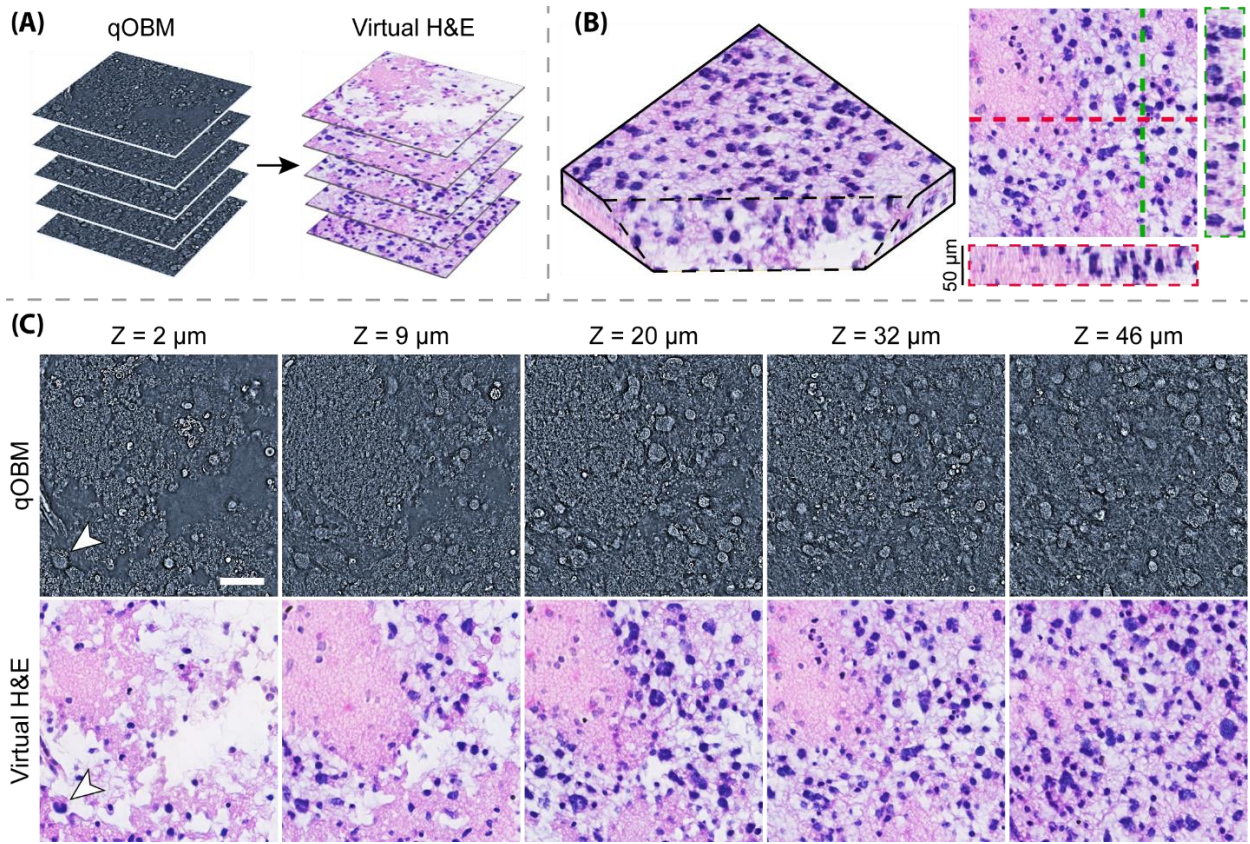

**Figure 9 - A qOBM and corresponding vH&E 3D volumetric stack of a human glioma margin.** A virtual H&E volumetric stack obtained by applying the trained human glioma CycleGAN to each image in the stack. **(A)** qOBM-to-vH&E conversion of the volumetric stack is depicted. **(B)** vH&E volume, with X-Y, X-Z, and Y-Z cross sections shown. **(C)** qOBM image slices at various depths and the corresponding vH&E image slices. The white arrow highlights an irregular nucleus. Scale bar is 50 µm. See **Supplementary Videos 3,4** for a depth sweep-through of this volumetric stack.

## Data Availability
Data required to replicate the experimental results is available at
https://github.com/tmabraham/qOBMtoHE.

## Code Availability
All of the code needed to replicate the experimental is provided at
https://github.com/tmabraham/qOBMtoHE.

## Acknowledgements
The authors would like to thank neuropathologist Dr. Stewart Neill, Dr. Bret Mobley, Dr. Joshua Klonoski, Dr. Jian Yi Li, and Dr. Viharkumar Patel for their participation in the vH&E review panel study. This work was funded by NIH R01EB028635, Stability AI PhD Fellowship, Burroughs Wellcome Fund (1014540); Marcus Center for Therapeutic Cell Characterization and Manufacturing (MC3M); National Cancer Institute (R21CA223853, R33CA202881); National Institute of General Medical Sciences (35GM147437); National Institute of Neurological Disorders and Stroke (R21NS117067); National Science Foundation (CAREER 1752011; GRFP DGE-2039655); Georgia Institute of Technology. This work was enabled by computing resources provided by the Wicklow AI in Medicine Research Initiative, the 2021 Spell Research Grant, and the Stability AI Academic Research Hardware Grant.

## Author Contributions
T.M.A. implemented, trained, and applied all virtual staining models and the classifier test. P.C.C. imaged rat brain specimens with qOBM and brightfield H&E. P.C.C. and C.F. imaged human brain specimens with qOBM and brightfield H&E. P.C.C. and C.F. prepared, matched, and labeled the data. C.F. conducted the pathologist validation study. Z.G. imaged mouse liver specimens with qOBM and H&E. Z.Z. and J.J.O provided the 9L rat tumor tissue and human surgical specimens. S.N. provided clinical interpretation of images. T.M.A., P.C.C., C.F., F.E.R. and R.L. wrote the manuscript. R.L. and F.E.R. conceived and supervised the research. All authors reviewed the manuscript.

## Competing Interests
R.L. is a co-founder of HistoliX, Inc. T.M.A. is a paid employee of Stability AI. The other authors declare that they have no competing financial interests.

## References
1. Bancroft, J. D. & Gamble, M. *Theory and Practice of Histological Techniques*. (Churchill Livingstone, 2008). doi:https://doi.org/10.1016/B978-0-443-10279-0.50001-4.
2. Fereidouni, F. *et al.* Microscopy with ultraviolet surface excitation for rapid slide-free histology. *Nature Biomedical Engineering* **1**, 957–966 (2017).
3. Glaser, A. K. *et al.* Light-sheet microscopy for slide-free non-destructive pathology of large clinical specimens. *Nature Biomedical Engineering* **1**, 1–10 (2017).
4. Tao, Y. K. *et al.* Assessment of breast pathologies using nonlinear microscopy. *PNAS* **111**, 15304–15309 (2014).
5. Jamme, F. *et al.* Deep UV autofluorescence microscopy for cell biology and tissue histology. *Biology of the Cell* **105**, 277–288 (2013).
6. Patel, K. B. *et al.* High-speed light-sheet microscopy for the in-situ acquisition of volumetric histological images of living tissue. *Nat. Biomed. Eng* **6**, 569–583 (2022).
7. Ye, S. *et al.* Rapid and label-free histological imaging of unprocessed surgical tissues via dark-field reflectance ultraviolet microscopy. *iScience* **26**, (2023).

8. Tu, H. *et al.* Stain-free histopathology by programmable supercontinuum pulses. *Nature Photonics* **10**, 534–540 (2016).
9. Witte, S. *et al.* Label-free live brain imaging with third-harmonic generation microscopy. in *CLEO/Europe and EQEC 2011 Conference Digest (2011), paper CLEB1_1* CLEB1_1 (Optica Publishing Group, 2011).
10. Ji, M. *et al.* Rapid, label-free detection of brain tumors with stimulated Raman scattering microscopy. *Sci Transl Med* **5**, 201ra119 (2013).
11. Orringer, D. A. *et al.* Rapid intraoperative histology of unprocessed surgical specimens via fibre-laser-based stimulated Raman scattering microscopy. *Nature Biomedical Engineering* **1**, 1–13 (2017).
12. Giacomelli, M. G. *et al.* Virtual Hematoxylin and Eosin Transillumination Microscopy Using Epi-Fluorescence Imaging. *PLOS ONE* **11**, e0159337 (2016).
13. Bai, B. *et al.* Deep learning-enabled virtual histological staining of biological samples. *Light Sci Appl* **12**, 57 (2023).
14. Rivenson, Y. *et al.* PhaseStain: the digital staining of label-free quantitative phase microscopy images using deep learning. *Light Sci Appl* **8**, 1–11 (2019).
15. Ledwig, P. & Robles, F. E. Epi-mode tomographic quantitative phase imaging in thick scattering samples. *Biomed Opt Express* **10**, 3605–3621 (2019).
16. Ledwig, P. & Robles, F. E. Quantitative 3D refractive index tomography of opaque samples in epi-mode. *Optica, OPTICA* **8**, 6–14 (2021).
17. Costa, P. C. *et al.* Towards in-vivo label-free detection of brain tumor margins with epi-illumination tomographic quantitative phase imaging. *Biomed. Opt. Express, BOE* **12**, 1621–1634 (2021).
18. Guang, Z., Ledwig, P., Costa, P. C., Filan, C. & Robles, F. E. Optimization of a flexible fiber-optic probe for epi-mode quantitative phase imaging. *Opt. Express, OE* **30**, 17713–17729 (2022).
19. Goodfellow, I. *et al.* Generative Adversarial Nets. in *Advances in Neural Information Processing Systems 27* (eds. Ghahramani, Z., Welling, M., Cortes, C., Lawrence, N. D. & Weinberger, K. Q.) 2672–2680 (Curran Associates, Inc., 2014).
20. Li, J. *et al.* Biopsy-free in vivo virtual histology of skin using deep learning. *Light Sci Appl* **10**, 233 (2021).
21. Boktor, M. *et al.* Virtual histological staining of label-free total absorption photoacoustic remote sensing (TA-PARS). *Sci Rep* **12**, 10296 (2022).
22. Zhu, J.-Y., Park, T., Isola, P. & Efros, A. A. Unpaired Image-to-Image Translation Using Cycle-Consistent Adversarial Networks. in *2017 IEEE International Conference on Computer Vision (ICCV)* 2242–2251 (IEEE, 2017). doi:10.1109/ICCV.2017.244.
23. Combalia, M. *et al.* Digitally Stained Confocal Microscopy through Deep Learning. in *International Conference on Medical Imaging with Deep Learning* 121–129 (PMLR, 2019).
24. Abraham, T., Shaw, A., O'Connor, D., Todd, A. & Levenson, R. Slide-free MUSE Microscopy to H&E Histology Modality Conversion via Unpaired Image-to-Image Translation GAN Models. *arXiv:2008.08579 [cs, eess]* (2020).
25. Cao, R. *et al.* Label-free intraoperative histology of bone tissue via deep-learning-assisted ultraviolet photoacoustic microscopy. *Nat. Biomed. Eng* **7**, 124–134 (2023).
26. Zhao, S. *et al.* Intraoperative fluorescence-guided resection of high-grade malignant gliomas using 5-aminolevulinic acid-induced porphyrins: a systematic review and meta-analysis of prospective studies. *PLoS One* **8**, e63682 (2013).
27. Hadjipanayis, C. G., Widhalm, G. & Stummer, W. What is the Surgical Benefit of Utilizing 5-ALA for Fluorescence-Guided Surgery of Malignant Gliomas? *Neurosurgery* **77**, 663–673 (2015).
28. Tonn, J.-C. & Stummer, W. Fluorescence-guided resection of malignant gliomas using 5-aminolevulinic acid: practical use, risks, and pitfalls. *Clin Neurosurg* **55**, 20–26 (2008).
29. Valdés, P. A. *et al.* δ-aminolevulinic acid-induced protoporphyrin IX concentration correlates with histopathologic markers of malignancy in human gliomas: the need for quantitative fluorescence-guided resection to identify regions of increasing malignancy. *Neuro Oncol* **13**, 846–856 (2011).

30. Kairdolf, B. A. *et al.* Intraoperative Spectroscopy with Ultrahigh Sensitivity for Image-Guided Surgery of Malignant Brain Tumors. *Anal Chem* **88**, 858–867 (2016).
31. Stummer, W., Reulen, H. J., Novotny, A., Stepp, H. & Tonn, J. C. Fluorescence-guided resections of malignant gliomas--an overview. *Acta Neurochir Suppl* **88**, 9–12 (2003).
32. Winetraub, Y. *et al.* OCT2Hist: Non-Invasive Virtual Biopsy Using Optical Coherence Tomography. 2021.03.31.21254733 Preprint at https://doi.org/10.1101/2021.03.31.21254733 (2021).
33. Oquab, M., Bottou, L., Laptev, I. & Sivic, J. Learning and Transferring Mid-Level Image Representations using Convolutional Neural Networks. in 1717–1724 (2014).
34. Casteleiro Costa, P. Quantitative oblique back-illumination microscopy in the study of biomedical samples. (Georgia Institute of Technology, 2023).
35. Liu, J. T. C. *et al.* Harnessing non-destructive 3D pathology. *Nat Biomed Eng* **5**, 203–218 (2021).
36. Li, M. *et al.* GAN Compression: Efficient Architectures for Interactive Conditional GANs. in 5284–5294 (2020).
37. Campanella, G. *et al.* Clinical-grade computational pathology using weakly supervised deep learning on whole slide images. *Nat. Med.* **25**, 1301–1309 (2019).
38. Lu, M. Y. *et al.* Data-efficient and weakly supervised computational pathology on whole-slide images. *Nature Biomedical Engineering* 1–16 (2021) doi:10.1038/s41551-020-00682-w.
39. Mertz, J. Optical sectioning microscopy with planar or structured illumination. *Nat Methods* **8**, 811–819 (2011).
40. Isola, P., Zhu, J.-Y., Zhou, T. & Efros, A. A. Image-to-Image Translation with Conditional Adversarial Networks. in *2017 IEEE Conference on Computer Vision and Pattern Recognition (CVPR)* 5967–5976 (2017). doi:10.1109/CVPR.2017.632.
41. He, K., Zhang, X., Ren, S. & Sun, J. Deep Residual Learning for Image Recognition. in *2016 IEEE Conference on Computer Vision and Pattern Recognition (CVPR)* 770–778 (2016). doi:10.1109/CVPR.2016.90.
42. Howard, J. & Gugger, S. *Deep Learning for Coders with fastai and PyTorch*. (O'Reilly Media, Inc., 2020).
43. Smith, L. N. A disciplined approach to neural network hyper-parameters: Part 1 -- learning rate, batch size, momentum, and weight decay. Preprint at https://doi.org/10.48550/arXiv.1803.09820 (2018).
44. Paszke, A. *et al.* PyTorch: An Imperative Style, High-Performance Deep Learning Library. in *Advances in Neural Information Processing Systems 32* (eds. Wallach, H. et al.) 8026–8037 (Curran Associates, Inc., 2019).
45. Howard, J. & Gugger, S. Fastai: A Layered API for Deep Learning. *Information* **11**, 108 (2020).
46. Abraham, Tanishq Mathew. UPIT - A fastai/PyTorch package for unpaired image-to-image translation. (2021) doi:10.5281/ZENODO.7889405.

# Supplementary Information

**Supplementary Note 1: Intensity inversion required for optimal CycleGAN conversion**

We found it necessary to invert the grayscale values of the native qOBM images, in order to render the nuclei to appear dark against a lighter background (as they typically appear on H&E). If this step is not taken, the nuclei in qOBM are frequently converted by the CycleGAN into white areas in vH&E (green arrows; **Supp Fig. 1**) while the darker cytoplasmic regions in qOBM are rendered erroneously as nuclei (blue arrows; **Supp Fig. 1**). A similar phenomenon was observed for virtual re-staining of MUSE, another microscopy modality, in Abraham et al.[1]

**Supplementary Note 2: Effect of model size on CycleGAN performance**

We note that CycleGAN model size is an important property that affects conversion quality. **Supplementary Figure 2** compares the conversions with a small model size/capacity (three layers in the discriminators, nine residual blocks in the generators) and large model size (six layers in the discriminators, twelve residual blocks in the generators) to the conversions of four separate CycleGANs with a small model capacity trained on images of each of the four observed subtypes. These results indicate that the small model size can capture histological features of individual tissue subtypes, but was unable to do so for multiple combined subtypes. Instead, as determined by visual examination, a larger model capacity was necessary to obtain conversions that capture the diversity of histological features seen in the multiple tissue subtypes.

**Supplementary Note 3: Additional comments on the performance of the classifier developed to differentiate tumors from healthy tissue.**

The classifier trained on H&E images from the 9L gliosarcoma tumor model to differentiate tumors from healthy tissue shows excellent accuracy when tested on vH&E images ($95.2 \pm 2.8\%$). Nevertheless, it is interesting to note which image types were misclassified. As the confusion matrix of the classifier applied to vH&E images shows (**Fig. 5B**), the misclassified regions are false positives, indicating that the CycleGAN conversion occasionally imparted tumor-like features to healthy images. As shown in **Supplementary Figure 4**, the misclassified regions are primarily a result of the CycleGAN hallucinating dark (hyperchromatic) tumor nuclei around blood vessels in otherwise acellular cortex region (**Supp Fig. 4A-B**) and basal ganglia (**Supp Fig. 4C-D**). Similar to the liver results presented above, this suggest that the network at times struggled to interpret structures around blood vessels which had a different appearance in the fresh tissues compared to the processed tissues. Again, it is likely that such failures can be mitigated with improved training of the qOBM-to-vH&E CycleGAN, using larger datasets or alternatively, unpaired image-to-image translation algorithms.

**Supplementary Note 4: Generalization of CycleGAN trained on rat to human specimens**

To examine the generalization of the qOBM-to-vH&E conversion, we took a neural network trained on rat images and applied it to the qOBM images of human specimens (**Supp Fig. 5A-B**). We observed that nuclear detail remained present, but the CycleGAN had a tendency to render nuclei as red blood cells, inappropriately convert whitespaces, and hallucinate other details (**Supp Fig. 5C-D**). This motivated the use of transfer learning to improve conversion efforts on the human glioma images.

**Supplementary Figures**

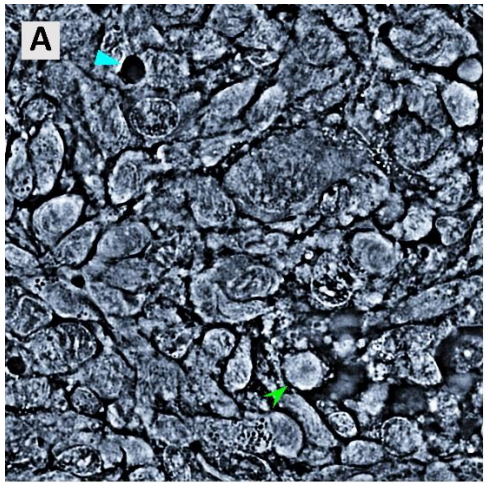
qOBM image

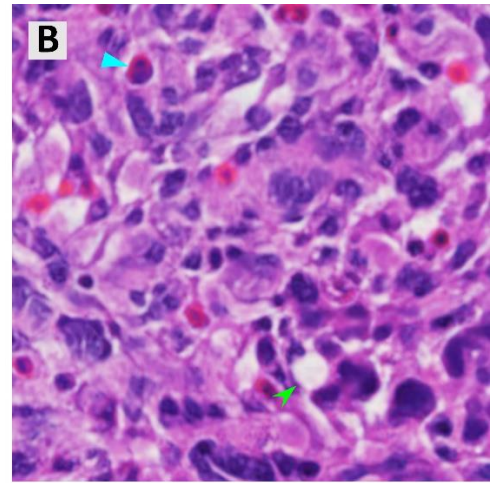
vH&E image from original qOBM

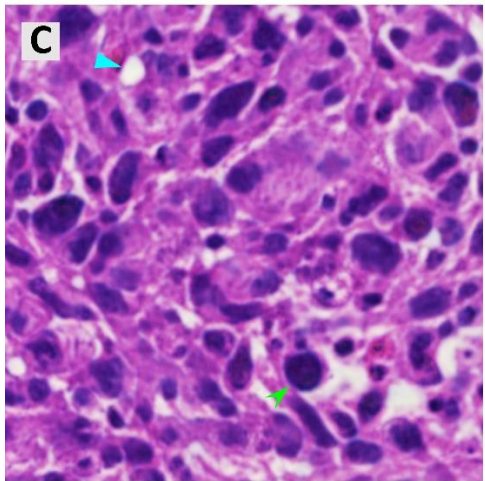
vH&E image from inverted qOBM

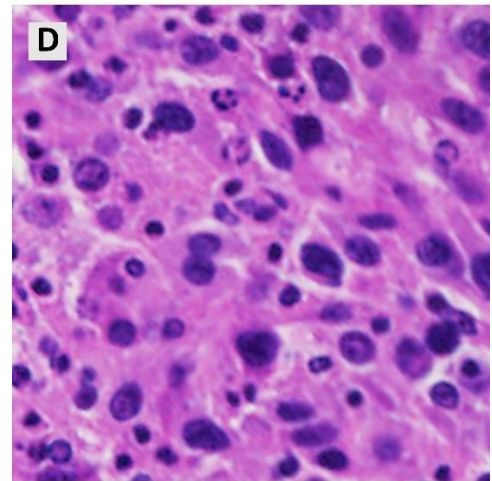
Standard brightfield H&E for comparison

**Supplementary Figure 1 - qOBM-to-vH&E conversion with original qOBM images. A** Label-free 60x qOBM image of a rat brain tumor region with spherical epithelioid cells. **B** vH&E produced by CycleGAN trained with the original qOBM images of the rat brain tumor regions with spherical epithelioid cells. **C** vH&E produced by CycleGAN trained with the contrast-inverted qOBM images of the rat brain tumor regions with spherical epithelioid cells. **D** Standard brightfield H&E image provided for comparison. Arrows highlight specific examples of inaccuracy during conversion.

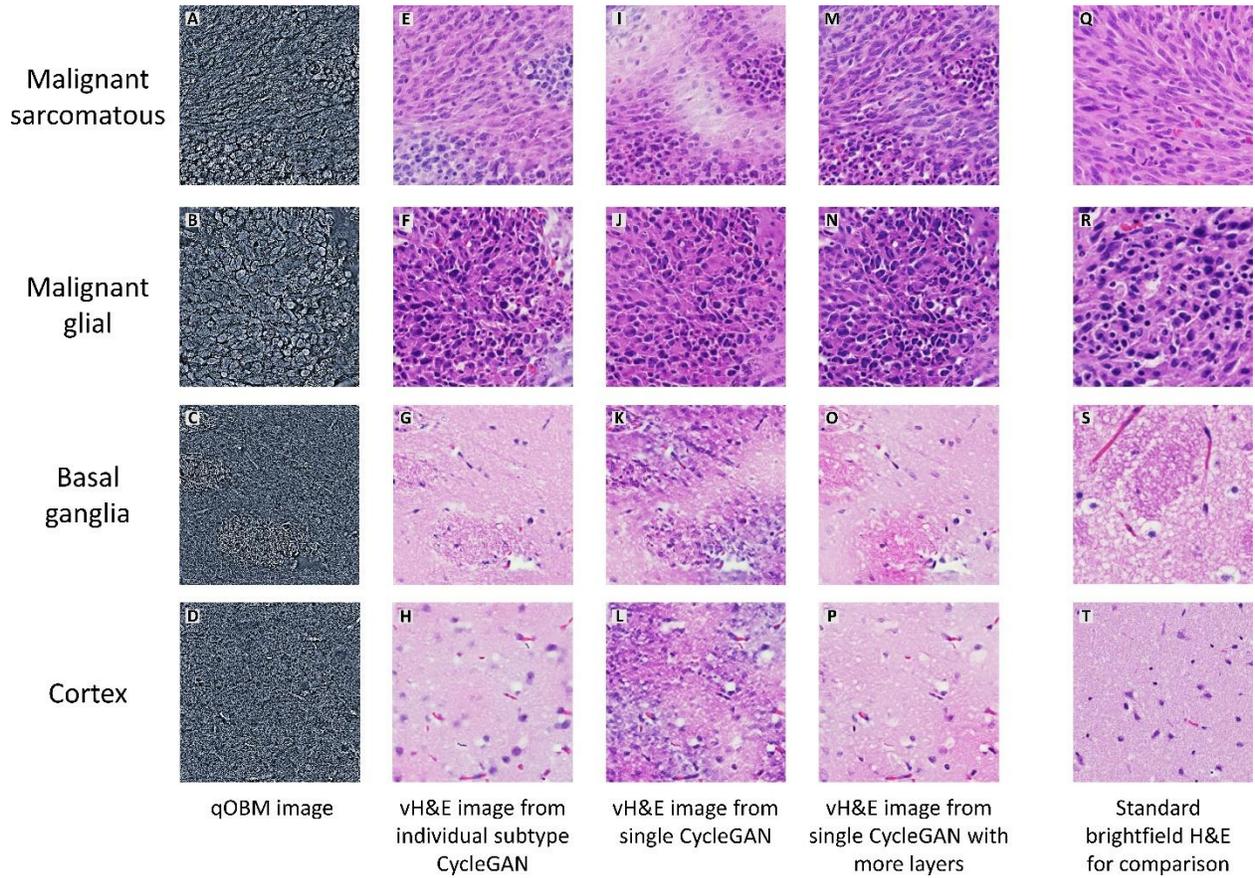

**Supplementary Figure 2 - qOBM-to-vH&E conversion on brain tissue subtypes using CycleGANs with different training setups. A-D** Label-free 60x qOBM images of each of the four rat brain tissue subtypes. **E-H** Corresponding vH&E images produced by separate CycleGANs for each of the four rat brain tissue subtypes. **I-L** Corresponding vH&E images produced by a single CycleGAN trained on all four rat brain tissue subtypes simultaneously. **M-P** Corresponding vH&E images produced by a single CycleGAN with more layers, trained on all four rat brain tissue subtypes simultaneously. **Q-T** Standard brightfield H&E images of the same tissue subtypes, provided for comparison.

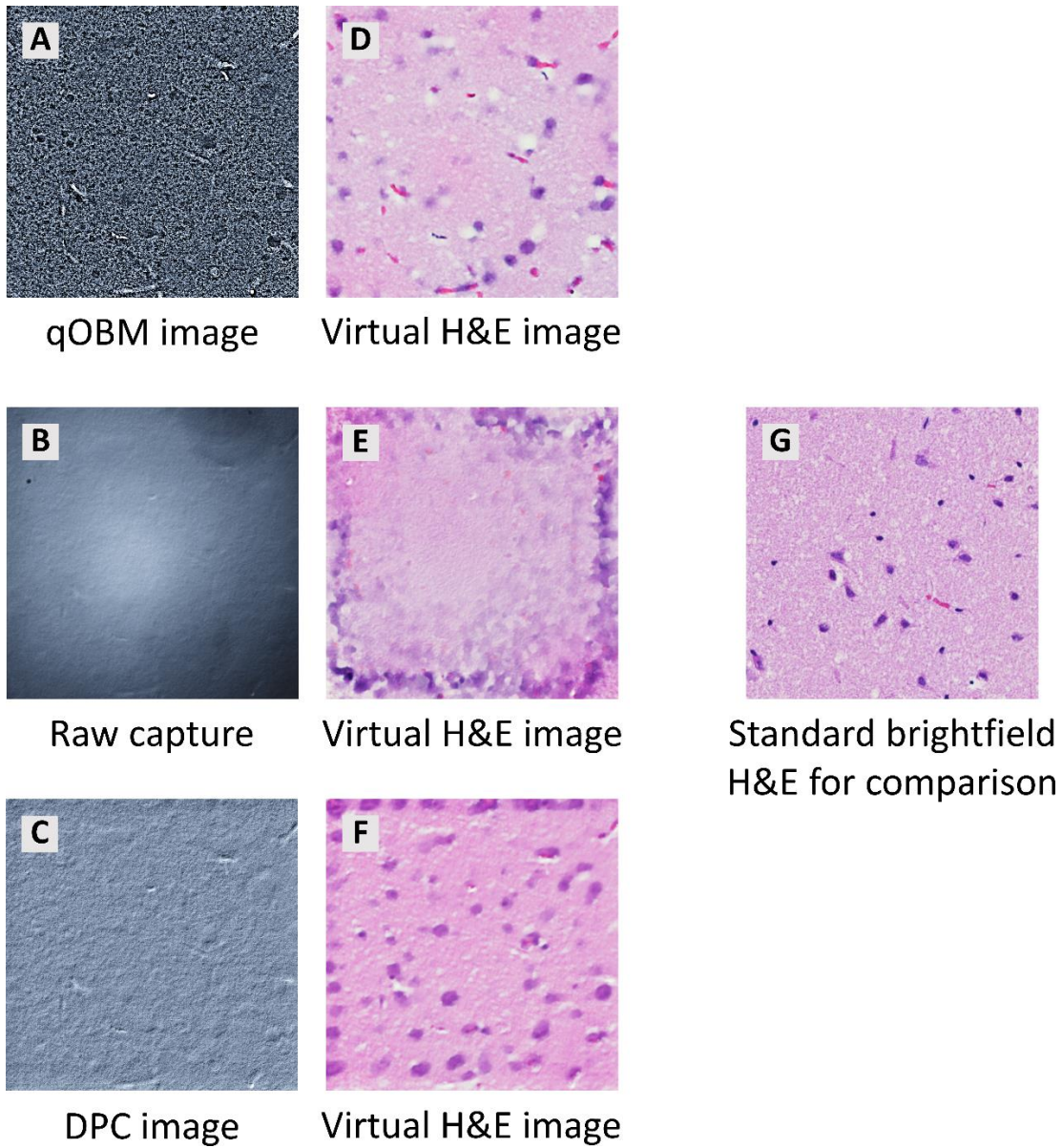

**Supplementary Figure 3** - Example vH&E conversions by CycleGANs trained on single capture, DPC, and qOBM images. **A** qOBM image of rat brain cortex. **B** One of the raw captures from the same field of view used to reconstruct the qOBM image. **C** One of the DPC images from the same field of view used to reconstruct the qOBM image. **D** vH&E image produced by a CycleGAN trained on rat brain cortex qOBM and H&E images. **E** vH&E image produced by a CycleGAN trained on rat brain cortex single capture and H&E images. **F** vH&E image produced by a CycleGAN trained on a rat brain cortex DPC and H&E images. **G** Standard brightfield H&E image provided for comparison.

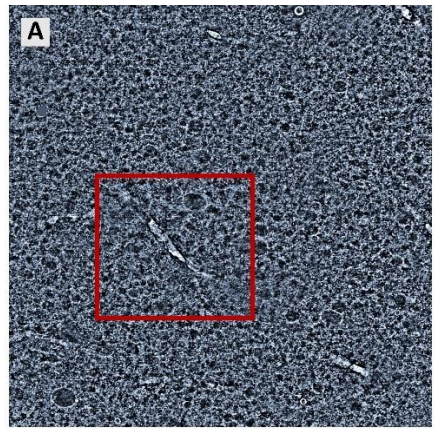
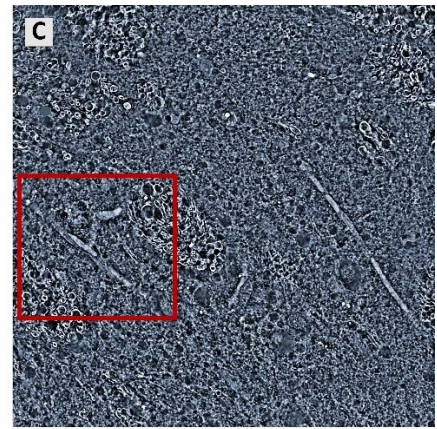

qOBM image

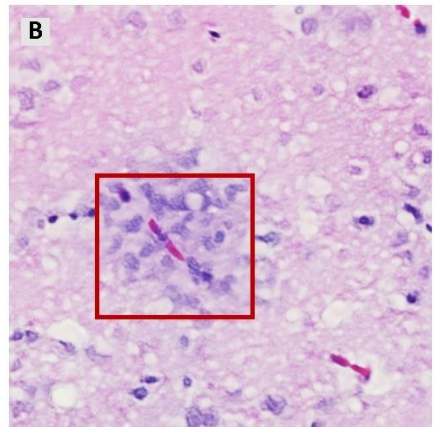
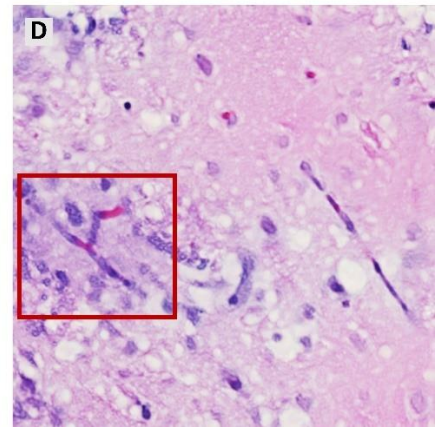

Virtual H&E image

**Supplementary Figure 4** - **Examples of poor qOBM-to-H&E conversions by the CycleGAN trained on rat brain images.** **A** 60x qOBM image of rat cortex. **B** Corresponding vH&E image. **C** 60x qOBM image from rat basal ganglia. **D** Corresponding vH&E image. Red boxes highlight location of significant conversion error. Conversion errors are primarily observed around vessels which are structures that deviate between the qOBM images of fresh tissues and H&E images of processed tissue sections. In qOBM images of fresh tissues, capillaries are better preserved and continuous; in comparison, capillaries appear fragmented in H&E images of processed tissue sections.

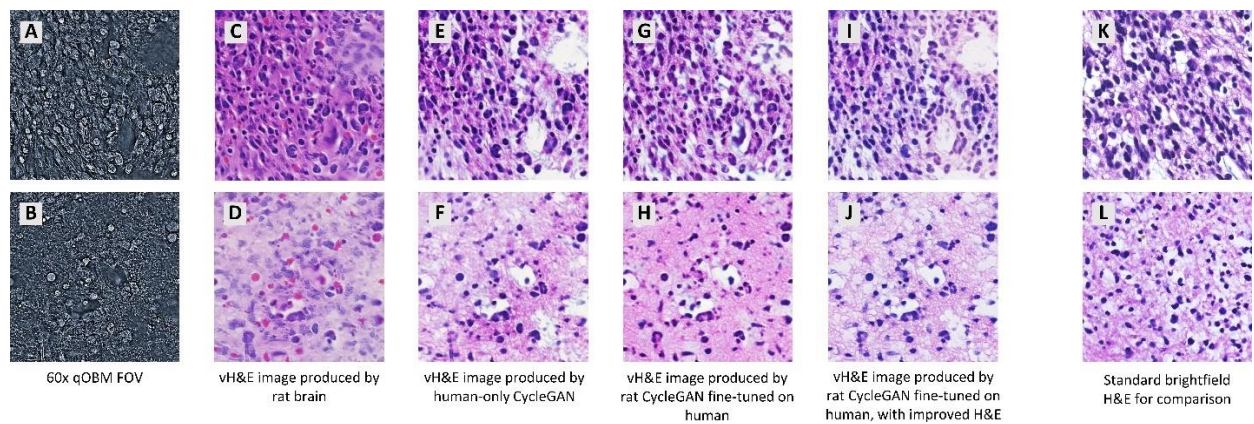

**Supplementary Figure 5 - qOBM-to-vH&E conversion of human glioma specimens using CycleGANs with different training and dataset setups**. **A-B** Label-free 60x qOBM images of human glioma specimens. **C-D** Corresponding vH&E images produced by the CycleGAN trained on only rat brain images. **E-F** Corresponding vH&E images produced by a CycleGAN trained only on human glioma images. **G-H** Corresponding vH&E images produced by a CycleGAN originally trained on rat brain images and further fine-tuned on human glioma images. **I-J** Corresponding vH&E images produced by a CycleGAN originally trained on rat brain images and further fine-tuned on human glioma images, using higher resolution H&E images. **K-L** Standard brightfield H&E images of the same tissue subtypes, provided for comparison.

# Supplementary References

1. Abraham, T., Shaw, A., O'Connor, D., Todd, A. & Levenson, R. Slide-free MUSE Microscopy to H&E Histology Modality Conversion via Unpaired Image-to-Image Translation GAN Models. *ArXiv200808579 Cs Eess* (2020).